\documentstyle[12pt]{article}
\begin{document}
\def\singlespace {\smallskipamount=3.75pt plus1pt minus1pt
                  \medskipamount=7.5pt plus2pt minus2pt
                  \bigskipamount=15pt plus4pt minus4pt
                  \normalbaselineskip=15pt plus0pt minus0pt
                  \normallineskip=1pt
                  \normallineskiplimit=0pt
                  \jot=3.75pt
                  {\def\smallskip {\vskip\smallskipamount}}
                  {\def\medskip   {\vskip\medskipamount}}
                  {\def\bigskip   {\vskip\bigskipamount}}
                  {\setbox\strutbox=\hbox{\vrule
                    height10.5pt depth4.5pt width 0pt}}
                  \parskip 7.5pt
                  \normalbaselines}
\begin{center}
{\bf {\Large Quantum General Relativity and}}
\smallskip
{\bf {\Large Hawking Radiation}}

\bigskip
\bigskip

{{Cenalo Vaz$^{a}$\footnote{e-mail address: cvaz@ualg.pt},
Claus Kiefer$^{b}$\footnote{e-mail address: kiefer@thp.uni-koeln.de},
T. P. Singh$^{c}$\footnote{e-mail address: tpsingh@tifr.res.in}, and
Louis Witten$^{d}$\footnote{e-mail address: witten@physics.uc.edu}}}

\bigskip
\bigskip

{\it $^{a}$Faculdade de Ci\^encias e Tecnologia}\\
{\it Universidade do Algarve, Faro, Portugal}

\medskip

{\it $^{b}$Institut f\"ur Theoretische Physik, Universit\"at zu K\"oln}\\
{\it Z\"ulpicher Str. 77, 50937 K\"oln, Germany}

\medskip

{\it $^{c}$Tata Institute of Fundamental Research,}\\
{\it Homi Bhabha Road, Mumbai 400 005, India}

\medskip

{\it $^{d}$Department of Physics, University of Cincinnati}\\
{\it Cincinnati, Ohio 45221-0011, USA}
\end{center}

\bigskip
\bigskip

\begin{abstract}
In a previous paper we have set up the Wheeler-DeWitt equation 
which describes
the quantum general relativistic collapse of a spherical dust cloud. In the
present paper we specialize this equation to the case of matter perturbations
around a black hole, and show that in the WKB approximation, 
the wave-functional describes an eternal black hole in equilibrium with a 
thermal bath at Hawking temperature.
\end{abstract}

\newpage

\section{Introduction}

Quantum gravitational effects are expected to modify 
the nature of singularities
that arise as the end state of the classical gravitational collapse 
of a compact object.
A concrete analytical model of classical spherical collapse is 
the Lema\^\i tre-Tolman-Bondi (LTB)
dust solution of Einstein equations, which shows that the 
singularity forming in the
collapse could be either naked or covered, depending on the 
choice of initial conditions
\cite{sj,bsvwd}.

Treating this dust collapse model as a classical background, 
one can quantize a massless
scalar field on this space-time using standard techniques 
\cite{bsvw1,bsvw2,bsv}. When the
classical collapse ends in a black hole, 
the quantization of the scalar field yields
the emission of Hawking radiation from the black hole, as expected. However,
a strikingly different result is obtained when the scalar field 
is quantized on a classical
background which ends in a naked singularity. 
It turns out that during the period of validity
of the semi-classical approximation (curvatures should be less 
than Planck scale), the
collapsing cloud emits only about one Planck unit of energy \cite{sap}. 
Moreover, because the
back-reaction does not become important so long as gravity can 
be treated classically, it
follows that the future evolution of the star is governed by 
quantum gravitational effects and
it is impossible to say, from the semi-classical approximation, 
whether the star radiates
away its energy on a short time scale or settles down into a 
black hole state. This is
completely different from the black hole case where essentially the 
entire star evaporates
via Hawking radiation during the semi-classical phase.

A full understanding of the gravitational collapse, both in the 
naked and in the covered case,
requires the application of quantum gravity. 
Given our limited understanding of quantum gravity
at present, perhaps the theory which is currently most suited for 
addressing dynamical quantum
collapse and questions regarding its final state is 
canonical quantum general relativity.
Although limited in its ultimate scope as a theory of quantum gravity, 
the canonical theory
{\it can} meaningfully address the issue of singularities in 
minisuperspace and midisuperspace
models, so long as one can tackle questions relating 
to operator ordering and regularization,
and provide a notion of time evolution in the theory.

The minisuperspace quantization of a collapsing null dust 
{\em shell} was analyzed in \cite{haj}
where it was shown how the classical singularity can be avoided in 
the quantum theory. In this model, the avoidance is a direct
consequence of the unitary time evolution -- since the wave-function
vanishes at $r=0$ for early times, it does so at any time. 
As a consequence, an ingoing quantum shell develops into a
superposition of ingoing {\em and} outgoing shell. Such a scenario has
interesting physical features that
cannot be seen in a semi-classical approximation: 
no event horizon can form, and there
is neither an information loss nor a naked singularity.

A midisuperspace program to study the canonical 
quantum dynamics of the LTB dust collapse
has been developed during the last two years \cite{vws}, 
following earlier pioneering work by Kastrup and Thiemann \cite{KT} and
Kucha\v r \cite{kuc} on the quantization of the Schwarzschild geometry. 
Here one sets up a
canonical description of the collapse, using the dust proper time, 
the area radius and the mass
function of the cloud as canonical configuration variables. 
The evolution is recorded by the
dust proper time. One then develops the quantization via the 
momentum constraint and the
Wheeler-DeWitt equation, to which a solution for a general 
mass function has been found using an ad hoc
delta-function regularization used by DeWitt \cite{Witt}. 
We show below that this regularization scheme is
equivalent to the WKB approximation. 
(A similar analysis has also been carried out for null dust
\cite{vws2} although the constraints that are obtained in this 
case are linear and there is no need for regularization.)

The Schwarzschild black hole can be viewed as a LTB model with a constant
mass function. We have applied \cite{vw} the program described 
above to this particular
case and shown how the horizon area quantization and 
entropy of the eternal black hole can be
understood in terms of quantized shells of matter that are 
trapped inside the horizon.

The purpose of the present paper is to show that the WKB 
solution is able also to describe an eternal black hole
in equilibrium with a thermal bath at Hawking temperature,
if the mass function is chosen appropriately. We argue
that Hawking radiation may be understood as a combination 
of the WKB and the Born-Oppenheimer
approximations on the full quantum wave-functional. 
It reinforces our belief in the overall
consistency of the program and, in particular, 
suggests that our proposed choice of operator
ordering and our definition of the inner product on the 
Hilbert space of wave-functionals capture features of the full theory. 
As we will see below, the definition 
of the inner product enters crucially in the
calculation of the Hawking radiation.

In Section 2, we briefly recall key results from our previous 
paper on quantum dust collapse
\cite{vws}, leading up to the WKB solution of the Wheeler-DeWitt equation. 
In Section 3, this
solution is specialized to the case of matter around a black hole, 
and shown to describe Hawking radiation.

%%%%%%%%%%%%%%%%%%%%%%%%%%%%%%%%%%%%%%%%%%%%%%%%%%%%%%%%%%%%%%%%%%%
\section{Canonical quantization of dust collapse}

The spherical gravitational collapse of a dust cloud having energy density 
$\epsilon (\tau ,\rho )$
in an asymptotically flat space-time is described in 
comoving coordinates $(\tau ,\rho ,\theta ,\phi )$ by 
the LTB metric
\begin{equation}
\label{ltb}ds^2=d\tau ^2-\frac{\tilde{R}^2(\rho,\tau)}{1+f(\rho)}
d\rho ^2-R^2(\rho,\tau)d\Omega ^2,
\end{equation}
and the Einstein equations
\begin{equation}
\label{ee}\epsilon(\tau,\rho) =\frac {\tilde{F}}{R^2\tilde{R}}\ ,~~~~ 
R^{*}(\tau,\rho)=\pm
\sqrt{f+F/R}\ .
\end{equation}
Here, $F(\rho)$ is twice the mass to the interior of the coordinate $\rho$ and $R(\tau,\rho)$ is
the area radius of the shell labeled $\rho$ at the dust proper time $\tau $. A tilde and an asterisk
denote partial derivatives with respect to $\rho$ and $\tau$, respectively.
(Throughout this paper, the gravitational constant is set equal to one.)

The canonical dynamics of the collapsing cloud is described by embedding the spherically symmetric ADM 4-metric,
\begin{equation}
\label{adm}ds^2=N^2dt^2-L^2(dr-N^rdt)^2-R^2d\Omega ^2,
\end{equation}
in the LTB space-time (\ref{ltb}), and by casting the action for the Einstein-dust system in
canonical form. The phase space of non-rotating dust is described by the dust proper time, $\tau$,
and its conjugate momentum, whereas the gravitational phase space consists of the configuration
space variables $(R,L)$ and their conjugate momenta. Using a version of the canonical transformation
developed by Kucha\v r \cite{kuc}, 
the configuration variable $L$ is replaced by a new variable $F$ (the mass
function). In terms of the new chart $(\tau, R,F,P_\tau, P_R,P_F)$, the momentum and the Hamiltonian
constraints read \cite{vws}
\begin{equation}
\label{mom}H_r=\tau ^{\prime }P_\tau +R^{\prime }P_R+F^{\prime }P_F\approx
0\ ,
\end{equation}
\begin{equation}
\label{ham}H=P_\tau ^2+{\cal F}P_R^2-\frac{F^{\prime 2}}{4%
{\cal F}}\approx 0\ .
\end{equation}
Here, ${\cal F}\equiv 1-F/R$. 
The Hamiltonian constraint shows that on the effective configuration space
$(\tau ,R)$, the DeWitt super-metric is just $\mathrm{diag}(1,1/{\cal F)}$. This is a flat metric,
therefore a redefinition of the area coordinate according to
\begin{equation}
\label{rst}
R_{*}=\pm \int \frac{dR}{\sqrt{|{\cal F|}}}\ , 
\end{equation}
where the positive sign refers to the region exterior to the horizon ($R> F$) and the negative
sign to the region interior to the horizon ($R<F$), brings the super-metric to manifestly flat form.
In terms of the momentum $P_*$, conjugate to $R_*$, the Hamiltonian constraint reads
\begin{equation}
H=P_\tau ^2\pm P_*^2-\frac{F^{\prime 2}}{4 {\cal F}}\approx 0\ .
\end{equation}
Quantization is implemented by raising the momenta to operator status and requiring the physical
state, $\Psi [\tau ,R_*,F]$, to be annihilated by the constraints. In this way, the time evolution
of $\Psi[\tau,R_*,F]$ is determined by the Hamiltonian constraint,
\begin{equation}
\label{wd}\left[ \frac{\delta ^2}{\delta \tau ^2}\pm \frac{\delta ^2}{\delta
R_{*}^2}+\frac{F^{\prime 2}}{4{\cal F}}\right] \Psi (\tau ,R,F)=0
\end{equation}
(where the positive sign before the second term refers to the region outside the horizon, and the
negative sign to the region inside), while invariance under spatial diffeomorphisms is implemented by
the momentum constraint,
\begin{equation}
\left[ \tau ^{\prime }\frac \delta {\delta \tau }+R_{*}^{\prime }\frac
\delta {\delta R_{*}}+F^{\prime }\frac \delta {\delta F}\right] \Psi [\tau
,R,F]=0\ . \label{mc}
\end{equation}
In the region exterior to the horizon, Eq.~(\ref{wd}) is no longer
hyperbolic, in contrast to the Wheeler-DeWitt equation on the
original configuration space. The reason lies in the canonical
transformations performed, which lead to a new effective configuration space.

To complete the quantum theory, one must define an inner product on the Hilbert space of
wave-functionals. In \cite{vws}, we defined it in a natural way by exploiting the fact that
the DeWitt super-metric is manifestly flat in the configuration space $(\tau,R_*)$,
\begin{equation}
\label{ip}\left\langle \Psi _{1}\vert\Psi _2\right\rangle
=\int_{R_{*}(0)}^\infty dR_{*}\bar{\Psi}_1\Psi _2\ .
\end{equation}
Note that this inner product is defined on a $\tau =$ constant hypersurface.
We emphasize that this inner product is in general $\tau$-dependent.
The reason is that the Wheeler-DeWitt equation preserves a
Klein-Gordon type of inner product, not a Schr\"odinger-type of
product \cite{Witt}. However, as has been shown in \cite{ks}, the
Schr\"odinger inner product is approximately conserved in the
highest orders of a semiclassical approximation. Since we shall deal here
with WKB states only (Sec.~3), quantum-gravitational correction terms
to this conservation do not play any role here.
Equations (\ref{wd})--(\ref{ip}) clearly imply a specific choice, albeit a natural one, of operator
ordering. We will see below that this choice is sufficient to reproduce the Hawking effect.

The momentum constraint is obeyed by {\it any} functional that is a spatial scalar and, in particular,
by the functional
\begin{equation}
\label{ss}\Psi [\tau ,R,F]=\exp \left[\frac 12\int_0^\infty drF^{\prime }(r){\cal %
W}(\tau (r),R(r),F(r))\right]\ ,
\end{equation}
provided that $\cal W$ has no explicit dependence on the radial label coordinate, $r$. Our choice,
while not unique, is dictated by the knowledge that $F'$ is the proper energy density of the collapsing
cloud. When (\ref{ss}) is substituted in the Wheeler-DeWitt equation (\ref{wd})
one finds, on using DeWitt's
$\delta-$function regularization ($\delta(0) = 0 = \delta^{(n)}(0)$ $\forall~ n \in \mathbf{N}$), that
$\cal W$ obeys
\begin{equation}
\label{wdw}\frac{F^{\prime 2}}4\left[ \left( \frac{\partial {\cal W}}{%
\partial \tau }\right) ^2\pm \left( \frac{\partial {\cal W}}{\partial R_{*}}%
\right) ^2+\frac 1{{\cal F}}\right] \Psi =0\ .
\end{equation}
We emphasize again that this regularization prescription is
at this stage completely ad hoc, and could only be justified from
an understanding of the full theory. It is even imaginable that,
analogous to string theory, Schwinger terms may arise in the commutation
relations of the constraints that could forbid the implementation of
the Wheeler-DeWitt equation \cite{CJZ}. Fortunately, however,
for our present purpose of recovering Hawking radiation it is not
necessary to resolve this issue.
 
Equation (\ref{wdw}) yields the solution
\begin{equation}
\label{out}{\cal W}=-i\tau \pm 2i\sqrt{F}\left[ \sqrt{R}-\sqrt{F}\tanh ^{-1}%
\sqrt{\frac FR}\right]
\end{equation}
outside the horizon, and
\begin{equation}
\label{in}{\cal W}=-i\tau \mp 2i\sqrt{F}\left[ \sqrt{R}-\sqrt{F}\tanh ^{-1}%
\sqrt{\frac RF}\right]
\end{equation}
inside.
%%%%%%%%%%%%%%%%%%%%%%%%%%%%%%%%%%%%%%%%%%%%%%%%%%%%%%%%%%%%%%%%%%%%%%%%

\section{Origin of Hawking radiation}

We begin by noting that the quantum constraint (\ref{wdw}), 
which has been written using DeWitt's
delta-function regularization, is the same equation as one 
would get by writing the Wheeler-DeWitt
equation (\ref{wd}) in the highest-order WKB approximation. The reason is that
this prescription effectively suppresses the (divergent) WKB prefactor.
 To show this, let us expand the wave-functional
$\Psi$ of (\ref{wd}) (with $\hbar$ being re-inserted)
 in a power-series in $\hbar$,
\begin{equation}
\label{app}\Psi \equiv \exp (iS/\hbar )\equiv \exp 
\left[\frac{1}{2\hbar}\int_0^\infty
drF^{\prime }(r){\cal S}(\tau (r),R(r),F(r))\right],
\end{equation}
\begin{equation}
\label{ess}{\cal S}(\tau ,R,F)={\cal S}_0+\hbar {\cal S}_1
+\hbar ^2{\cal S}_2+\ldots\ .
\end{equation}
Substituting this expansion in (\ref{wd}) and 
retaining only the leading order, $\hbar $-independent, terms gives
\begin{equation}
\label{wkb}\frac{F^{\prime 2}}4\left[ \left( \frac{\partial {\cal S}_0}{%
\partial \tau }\right) ^2\pm \left( \frac{\partial {\cal S}_0}{\partial R_{*}%
}\right) ^2-\frac 1{{\cal F}}\right] \Psi =0\ .
\end{equation}
Comparison with (\ref{wdw}) shows that the WKB solution is 
the same as one would get by doing the
delta-function regularization in the original Wheeler-DeWitt equation (after the identification
${\cal W}=i{\cal S}_0$). 
(For a similar discussion of WKB states for the Schwarzschild black hole
see \cite{brk} and for two-dimensional dilaton gravity see
\cite{kun}).

We will now show that the wave-functional (\ref{ss}), along with the solutions (\ref{out}) and
(\ref{in}), yields Hawking radiation when it is applied to a matter distribution that is
appropriate to a massive black hole surrounded by dust whose total energy is small compared
with the mass of the black hole. For this purpose, let us assume that the mass function $F(r)$ is of the form
\begin{equation}
\label{mas}F(r)=2M\theta (r)+f(r)\ ,
\end{equation}
where $\theta (r)$ is the Heaviside step-function, 
and $f(r)$ (not to be confused with $f(\rho)$ occurring
in (\ref{ltb})) is differentiable, representing
a dust distribution with $f(r)/2M\ll 1.$ This mass function is interpreted as the presence of a
Schwarzschild black hole of mass $M$ at the origin, and $f(r)$ is a dust matter perturbation on
the black hole, which, as we now show, can be related to Hawking radiation.
In a sense, it plays the role of the quantum field used in standard
derivations of Hawking radiation.

Inserting this mass function in the wave-functional (\ref{ss}) gives
(setting $\hbar=1$ again)
\begin{equation}
\label{expa}\Psi [\tau ,R,F]=\exp \left[\frac 12\int_0^\infty dr[2M\delta
(r)+f^{\prime }(r)]{\cal W}(\tau (r),R(r),F(r))\right]
\end{equation}
or
\begin{equation}
\label{exp2}\Psi [\tau ,R,F]\equiv
\exp \left[M{\cal W}_0(\tau ,R,F)\right]\times \exp \left[\frac
12\int_0^\infty drf^{\prime }(r){\cal W}^f(\tau (r),R(r),F(r))\right]
\end{equation}
where ${\cal W}_0(\tau ,R,F)\equiv
{\cal W}(\tau(0),R(0),F(0))$.
 The first exponent on the right-hand side is the WKB wave-functional
representing the black hole at the origin, as shown in \cite{vw}. The second term, up to order
$f(r)$, represents a matter distribution that propagates in this background, if we take $F(r)
\approx 2M$ in ${\cal W}^f$. Thus we have
\begin{equation}
\label{spl}\Psi =\Psi_{\rm bh}\times \Psi_f\ ,
\end{equation}
where
\begin{eqnarray}
\nonumber
\Psi_{\rm bh} &\sim& \exp \left[M{\cal W}_0(\tau ,R,F)\right]\ ,\cr\cr
\Psi_f &\sim& \exp \left[\frac 12\int_0^\infty drf^{\prime }(r){\cal W}^f(\tau (r),R(r),F(r))\right]\ .
\end{eqnarray}
The solution $\Psi _f$ is known from (\ref{ss}), (\ref{out}) and (\ref{in}) above and given by
\begin{equation}
\label{mo}{\cal W}_{\rm out}^f=-i\left[ \tau \mp 2\sqrt{2M}\left( 
 \sqrt{R}-\frac{\sqrt{2M}}2\ln \left[ \frac{\sqrt{R}+\sqrt{2M}}{\sqrt{R}-\sqrt{2M}}\right]\right) \right]
\end{equation}
outside the horizon, and
\begin{equation}
\label{mi}{\cal W}_{\rm in}^f=-i\left[ \tau \pm
 2\sqrt{2M}\left( \sqrt{R}-
\frac{\sqrt{2M}}2\ln \left[ \frac{\sqrt{R}+\sqrt{2M}}
{\sqrt{2M}-\sqrt{R}}\right]
\right) \right]
\end{equation}
inside the horizon.

We would like to rewrite the expressions for ${\cal W}^f$ in terms of the Killing time $T$.
For the Schwarzschild background being considered for the distribution $f(r)$ and for contracting
clouds, we have the following relation between the proper time and the Killing time (see e.g. \cite{fp}),
\begin{eqnarray}
\tau& =&T+\sqrt{2M}\int dR\frac{\sqrt{R}}{R-2M}\\
\ &=&T+2\sqrt{2M}\left[ \sqrt{R}-\frac{\sqrt{2M}}2\ln \left( \frac{\sqrt{R}+%
\sqrt{2M}}{\sqrt{R}-\sqrt{2M}}\right) \right]\ .
\label{sch}
\end{eqnarray}
Thus, in terms of $T$, there are two possibilities for ${\cal W}_{\rm out}^f$,
\begin{equation}
{\cal W}^{f}_{\rm out}=\left\{\begin{array}{ll}
-iT\ , \\
-i\left[ T + 4\sqrt{2M}\left( \sqrt{R}-\frac{\sqrt{2M}}2\ln \left[ \frac{\sqrt{R}+
\sqrt{2M}}{\sqrt{R}-\sqrt{2M}}\right] \right) \right]\ .
\end{array}\right.
\end{equation}
The wave-functional that corresponds to an infalling wave at $T\rightarrow -\infty $ and
$R\rightarrow \infty $ is the one with ${\cal W}_{\rm out}^f$ 
given by the second of the above.
We will therefore concentrate on
\begin{equation}
\label{pf}\Psi _f=\exp \left[ -i\int drf^{\prime }(r)\left( T+4\sqrt{2M}%
\left( \sqrt{R}-\frac{\sqrt{2M}}2\ln \left[ \frac{\sqrt{R}+\sqrt{2M}}{\sqrt{%
R}-\sqrt{2M}}\right] \right) \right) \right] .
\end{equation}
Defining $Z=4\sqrt{2MR}$ we find that as $R\rightarrow \infty $ this wave-functional approaches
\begin{equation}
\label{ass}\Psi_f^{-}\approx
\exp \left[-i\int drf^{\prime }(r)[T+Z] \right]
\end{equation}
which undergoes rapid oscillations except when $T\rightarrow -\infty $, that is on ${\cal I}^{-}$. When
$T\rightarrow \infty $, in order for the phase to be not large we see that $R\rightarrow 2M$,
i.e. $Z\rightarrow 8M$. In this limit we have
$$
\Psi_f^{+}=\exp\left\{ -i\int drf^{\prime }(r)\left( T+\left[ Z-4M\ln \left( \frac{
Z+8M}{Z-8M}\right) \right] \right)\right\}
$$
\begin{equation}
\label{asp}\approx \exp \left\{-i\int drf^{\prime }(r)\left( T+4M\ln \left( \frac{%
Z-8M}{16M}\right) \right) \right\}\ .
\end{equation}
This is similar to what happens in the geometric optics approximation. The simple looking
phase on ${\cal I}^{-}$ has scattered through the geometry to turn into the complicated
looking phase on ${\cal I}^{+}$ near the horizon.

Equation (\ref{ass}) represents infalling waves. We can think of it as a product over plane waves, one at each label $r$, as follows:
\begin{equation}
\label{pw}\Psi_\omega^{-}=\prod_re^{-i\omega (r)[T(r)+Z(r)]}\ .
\end{equation}
This should represent a complete set of infalling modes at each label $r$, if we think of
the $\omega (r)$ as the frequency of the modes. In other words, we allow all possible $\omega(r)=\Delta f(r)$. 

A complete set of outgoing modes on ${\cal I}^{+}$ would likewise be given by the functional
\begin{equation}
\label{og}{\Psi}_\omega^{+}=\prod_re^{-i\omega (r)[T(r)-Z(r)]}\ .
\end{equation}
This is because the transformation from the dust proper time to the Killing time is now
obtained by matching an expanding (rather than contracting) dust cloud to a Schwarzschild exterior,
for which one gets, instead of (\ref{sch}), the relation
\begin{eqnarray}
\label{sch2}\tau& =&T-\sqrt{2M}\int dR\frac{\sqrt{R}}{R-2M}\\
\ &=&T-2\sqrt{2M}\left[ \sqrt{R}-\frac{\sqrt{2M}}2\ln \left( \frac{\sqrt{R}+%
\sqrt{2M}}{\sqrt{R}-\sqrt{2M}}\right) \right].
\end{eqnarray}
It is then easily shown that as $R\rightarrow\infty$, the asymptotic form of the wave-functional is as in (\ref{og}).

Now we ask the question: what is the projection of our solution (\ref{asp}) on the negative
frequency modes of the outgoing basis on ${\cal I}^{+}$. For this purpose we must consider the inner product of states on a hypersurface of constant Schwarzschild time $T$. Thus we must transform from the Euclidean flat metric on the
$(R_{*},\tau)$ plane of Eqn. (\ref{wd}) to the metric in the $(R,T)$ coordinates. Using the relations (\ref{rst}) and (\ref{sch}) we get
$$
g_{RR}=\left({\partial R_{*}\over \partial R}\right)^{2}g_{R_{*}R_{*}} +
\left({\partial \tau\over \partial R}\right)^{2}g_{\tau\tau}=
\left({R\over R-2M}\right)^{2}\label{grr}
$$
(Note that in (\ref{wd}) the positive sign holds outside the horizon, so that
$g_{R_{*}R_{*}}=+1$.) The required inner product on a constant $T$ 
hypersurface then is
$$
\left\langle \Psi_\omega\vert\Psi _f^{+}\right\rangle =\prod_r\int
\sqrt{g_{RR}}dR(r)\Psi _\omega ^{+}(Z(r),T(r),\omega (r))\Psi _f^{+}(Z(r),T(r),f(r))
$$
$$
=\prod_r\int dR\frac R{R-2M}\Psi _\omega ^{+}(Z(r),T(r),\omega (r))\Psi
_f^{+}(Z(r),T(r),f(r))
$$
$$
=\prod_r\int dZ\frac{Z^3}{16M(Z^2-64M^2)}\Psi _\omega ^{+}(Z,T,\omega )
\Psi_f^{+}(Z,T,f)
$$
\begin{equation}
\label{mag}\approx \prod_r\int dZ\frac {2M}{Z-8M}\Psi _\omega ^{+}(Z,T,\omega
)\Psi _f^{+}(Z,T,f)\ .
\end{equation}
This projection represents the negative frequency modes present in the
solution. We are interested in $|\left\langle \Psi _\omega ^{+}\vert\Psi
_f^{+}\right\rangle |^2$ because these are the analogs of the Bogoliubov
coefficients, $|\beta (f,\omega )|^2.$ If we think of $%
\beta (f,\omega )$ as
\begin{equation}
\label{bet}\beta (f,\omega )=\prod_r\beta (f(r),\omega (r))\ ,
\end{equation}
we have%
$$
\beta (f(r),\omega (r))\approx 2M\int_{8M}^\infty \frac{dZ}{Z-8M}e^{i\omega
Z}e^{-4iM\Delta f\ln (\frac{Z-8M}{16M})}
$$
\begin{equation}
\label{be}=\frac{2M}{\left( 16M\right) ^{-4iM\Delta f}}\int_{8M}^\infty
dZ\left( Z-8M\right) ^{-1-4iM\Delta f}e^{i\omega Z}\ .
\end{equation}
Substituting $u=Z-8M$ and integrating we find%
$$
\beta (f(r),\omega (r))=\frac{2Me^{8Mi\omega }}{\left( 16M\right)
^{-4iM\Delta f}}\left[ \Gamma (-i\sigma )\omega ^{i\sigma }e^{-\sigma \pi
/2}\right]\ ,
$$
where $\sigma =4M\Delta f$. This gives
\begin{equation}
\label{haw}|\beta (f,\omega )|^2=\prod_r\frac{2\pi M}{\Delta f}\left[
\frac 1{e^{8\pi M\Delta f}-1}\right].
\end{equation}
This is interpreted as the eternal black hole being in equilibrium with a thermal bath at the Hawking temperature $(8\pi
M)^{-1}.$
Our derivation provides a functional Schr\"odinger picture for dust Hawking
radiation, consistent with the WKB wave-functional 
which solves the Wheeler-DeWitt equation.

%%%%%%%%%%%%%%%%%%%%%%%%%%%%%%%%%%%%%%%%%%%%%%%%%%%%%%%%%%%%%%%%%
\section{Concluding remarks}

In this paper we have obtained a derivation of Hawking radiation for dust
matter, starting from the WKB wave-functional which satisfies the
Wheeler-DeWitt equation for quantum spherical dust collapse. The fact that
such a derivation could be found should be treated as support for the
validity of the inner product defined in Equation (\ref{ip}).

A functional description of Hawking radiation can also be given 
within a Born-Oppenheimer type of approximation to quantum gravity.
Instead of our wave function $\Psi_f$, one has there a Gaussian
quantum state for a quantum field. Evolving this state through the
background of an object collapsing to a black hole, one finds that it
encodes information about the Hawking radiation similar to our
$\Psi_f$ \cite{dek}. 

The
present analysis also suggests that an exact treatment of the Wheeler-DeWitt
equation (\ref{wd}) which goes beyond the WKB approximation should yield
corrections to Hawking radiation, and provide a better understanding of the
end state of gravitational collapse. This could perhaps be done along the
general lines presented in \cite{ks} in which corrections to the
semiclassical limit have been calculated from the Wheeler-DeWitt equation. 
These issues are at present under investigation.

%%%%%%%%%%%%%%%%%%%%%%%%%%%%%%%%%%%%%%%%%%%%%%%%%%%%%%%%%%%%%%%%%%%%%%%
\section*{Acknowledgments}
The visit of T.P.S. to the University of Cologne was supported
by DFG grant KI 381/3-1. C.V, T.P.S. and L.W. acknowledge the
partial support of FCT, Portugal under contract SAPIENS/32694/99.
L.W. was supported in part by the
Department of Energy, USA, under Contract Number DOE-FG02-84ER40153.
%%%%%%%%%%%%%%%%%%%%%%%%%%%%%%%%%%%%%%%%%%%%%%%%%%%%%%%%%%%%%%%%


\begin{thebibliography}{99}

\bibitem{sj}
T. P. Singh and P. S. Joshi, Class. Quantum Grav. 13, 559 (1996).
\bibitem{bsvwd}
S. Barve, T. P. Singh, C. Vaz, and L. Witten, Class. Quantum Grav. 16, 1727
(1999).
\bibitem{bsvw1}
S. Barve, T. P. Singh, C. Vaz, and L. Witten, Nucl. Phys. B532, 361 (1998).
\bibitem{bsvw2}
S. Barve, T. P. Singh, C. Vaz, and L. Witten, Phys. Rev. D58, 104018 (1998).
\bibitem{bsv}
S. Barve, T. P. Singh, and C. Vaz, Phys. Rev. D62, 084021 (2000).
\bibitem{sap}
T. Harada, H. Iguchi, K. Nakao, T. P. Singh, T. Tanaka, and C. Vaz, Phys.
Rev. D64, 041501 (2001).
\bibitem{haj} P. H\'aj\'{\i}\v{c}ek and C. Kiefer, 
Nucl. Phys. B603, 531 (2001);
P. H\'aj\'{\i}\v{c}ek, ibid., 555 (2001); 
P. H\'aj\'{\i}\v{c}ek and C. Kiefer, Int. J. Mod. Phys. D10, 775 (2001);
P. H\'aj\'{\i}\v{c}ek, {\tt gr-qc/0204049}.
\bibitem{vws} C. Vaz, L. Witten, and T. P. Singh, Phys. Rev. D63, 
104020 (2001).
\bibitem{KT} H. A. Kastrup and T. Thiemann, Nucl. Phys. B425, 665 (1994).
\bibitem{kuc} K. V. Kucha\v{r}, Phys. Rev. D50, 3961 (1994).
\bibitem{Witt} B.S. DeWitt, Phys. Rev. 160, 1113 (1967).
\bibitem{vws2} C. Vaz, L. Witten, and T. P. Singh, 
Phys. Rev. D65, 104016 (2002).
\bibitem{vw} C. Vaz and L. Witten, Phys. Rev. D64, 084005 (2001).
\bibitem{ks} C. Kiefer and T. P. Singh, Phys. Rev. D44, 1067 (1991);
A. O. Barvinsky and C. Kiefer, Nucl. Phys. B526, 509 (1998).
\bibitem{CJZ} D. Cangemi, R. Jackiw, and B. Zwiebach,
 Ann. Phys. (N.Y.) 245, 408 (1996).
\bibitem{brk} T. Brotz and C. Kiefer, Phys. Rev. D55, 2186 (1997).
\bibitem{kun} D. Louis-Martinez, J. Gegenberg, and G. Kunstatter,
 Phys. Lett. B321, 193 (1994).
\bibitem{fp} L. H. Ford and L. Parker, Phys. Rev. D17, 1485 (1978).
\bibitem{dek} J.-G. Demers and C. Kiefer, Phys. Rev. D53, 7050 (1996).

\end{thebibliography}
\end{document}